# Systematic Task Allocation Evaluation in Distributed Software Development

Jürgen Münch[1], Ansgar Lamersdorf[2]

[1]Fraunhofer IESE, Fraunhofer Platz 1, 67663 Kaiserslautern, Germany
Juergen.Muench@iese.fraunhofer.de
[2]University of Kaiserslautern, Gottlieb-Daimler-Str., 67653 Kaiserslautern, Germany
a_lamers@informatik.uni-kl.de

Systematic task allocation to different development sites in global software development projects can open business and engineering perspectives and help to reduce risks and problems inherent in distributed development. Relying only on a single evaluation criterion such as development cost when distributing tasks to development sites has shown to be very risky and often does not lead to successful solutions in the long run. Task allocation in global software projects is challenging due to a multitude of impact factors and constraints. Systematic allocation decisions require the ability to evaluate and compare task allocation alternatives and to effectively establish customized task allocation practices in an organization. In this article, we present a customizable process for task allocation evaluation that is based on results from a systematic interview study with practitioners. In this process, the relevant criteria for evaluating task allocation alternatives are derived by applying principles from goal-oriented measurement. In addition, the customization of the process is demonstrated, related work and limitations are sketched, and an outlook on future work is given.

## 1. Introduction

Global Software Development (GSD) has become reality in many software development organizations, due to its promising benefits such as decreased labor costs and access to a worldwide pool of resources [1]. However, its inherent risks and complexity increase the difficulty and failure rate of GSD compared to single-site development [2].

The allocation of tasks, in particular (i.e., the decision on how to structure a GSD project and assign the work to different locations throughout the world), has a large impact on the success of distributed development projects and is influenced by several different criteria ([3], [4]). The authors hypothesize that, on the one hand, "smart globalization" (i.e., distributing work based upon systematic consideration of relevant criteria) can be the basis for many business and engineering prospects in GSD. On the other hand, omitting systematic evaluation of alternatives or having only one decision criterion (e.g., labor cost rates [5]) largely increases the risks of GSD. We thus see a need for the systematic selection and evaluation of task allocation alternatives.

This article presents an approach for systematically evaluating task allocations in GSD. As the criteria and factors influencing evaluation are very much dependent on the organization, we do not give a specific model but instead discuss a series of logical steps. They are based on principles from the Goal/Question/Metric (GQM) approach [6], a framework for the derivation of measures from goals (such as evaluation goals). GQM has been widely accepted and comes with a set of guidelines and sheets [7].

The article is structured as follows: First, the task allocation decision problem is explained in a specific scenario based on the results of an empirical study. Section 3 presents related work. The approach is presented together with its application on the given scenario. Finally, limitations and future work are sketched.

## 2. Scenario of a Task Allocation Decision Problem

In this section, we present a scenario for a typical task allocation problem in global software development in order to highlight important challenges and constraints. The scenario is based on the results of a qualitative study the authors conducted in order to analyze the state of the practice in task allocation [3]. In the following, we will briefly summarize the relevant findings of this study and then introduce the scenario, which will be used as an example throughout the remainder of this article.

***Empirical Basis for the Scenario***. We conducted a systematic interview study with 12 practitioners from different companies with several years of experience in distributed and global development [3]. The main goal of the study was to identify the industrial practice in task allocation, especially with respect to the criteria applied. Thus, the interviewees were asked to name the general background of distributed development at their company and to describe in detail the task allocation process and the applied criteria for one specific past project.

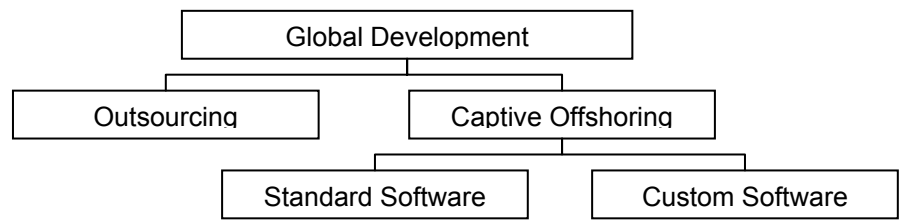

**Fig. 1.** Types of distributed development

The main result was that the strategy for task allocation very much depends on the type of distributed development (see Figure 1): While in software development outsourcing, large pieces of work (e.g., complete projects or products to be developed) are usually assigned to outside contractors, task assignment in captive offshoring (i.e., within one organization that has globally distributed sites) is done on a much finer level of granularity.

Captive offshoring can be further classified into development of standard software (e.g., shrink-wrapped software) and project-based development of custom software for individual clients. We found that in standard software development, assignment is largely done based on specialized teams that evolve over a long time. In custom software development, however, there is larger freedom in the assignment, and tasks are mostly allocated by availability of resources only.

The study also revealed the importance of cost rates as a driving force for GSD. Low labor cost rates were the most prominent criterion both for initiating global development and establishing new sites: New sites were built in low-cost regions in order to leverage labor costs. In custom development projects, there is also often pressure towards assigning more work to the low-cost sites, if possible

Out of the three identified types, we will focus on the development of custom software due to several reasons: On the one hand, the task allocation decision is highly complex, since there is a large degree of freedom in assigning work to sites and multiple influencing factors (e.g., cost rate, expertise, proximity to customer) have to be considered. On the other hand, in practice task assignment is typically unsystematic, based just on availability and cost rates. We thus see a high potential here for avoiding or reducing development risks as well as many opportunities for gaining benefits from distributed development.

Another finding of the study was that in many cases of custom software development, the requirements and the coarse architecture are derived at a central site, followed by the assignment of development work to the available sites. This means that task allocation in these cases is an assignment of the development of coarse architectural components to sites. In the following scenario, we will use this as a basis for the description of the tasks to be assigned.

*Task Allocation Scenario*. In order to illustrate the scenario, we introduce the "GlobalSoft" project as an exemplary case of custom software development in GSD. Even though it does not specifically reflect one of the projects described in the interview study, it is based on the experiences reported there.

GlobalSoft is a large, Europe-based software company that develops individual software products for customers in Germany and the UK. Its main development centers are located in Frankfurt and Cologne, Germany, and a smaller subsidiary also exists in London, UK, in order to have a site close to the British customers. Recently the company also established a site in Bangalore, India, in order to reduce labor costs and gain access to a large pool of software engineers. Figure 2 gives an overview of the available sites together with the labor cost rates per site.

In our scenario, GlobalSoft is developing new software for one of its customers, BigIndustries. BigIndustries is located in London and is well known to GlobalSoft due to a set of previously conducted projects. The old projects were always done at the sites in London, Frankfurt, and Cologne. In this new project, there is also the possibility of assigning work to the new development center in Bangalore.

At the time of the task allocation decision, requirements engineering and high-level architecture design have already been done at the sites in London and Frankfurt. Project management (located in Frankfurt) now has to decide how to assign the development of the identified architectural components to the sites. In addition, system testing and integration have to be assigned, too. Figure 3 shows the resulting tasks that are to

4      Jürgen Münch, Ansgar Lamersdorf

be assigned together with the expected effort distribution. As the high-level architecture already exists, it is also possible to estimate the expected coupling between the components. In the scenario, we assume that the coupling of the components is an indicator for the communication needed between sites [8].

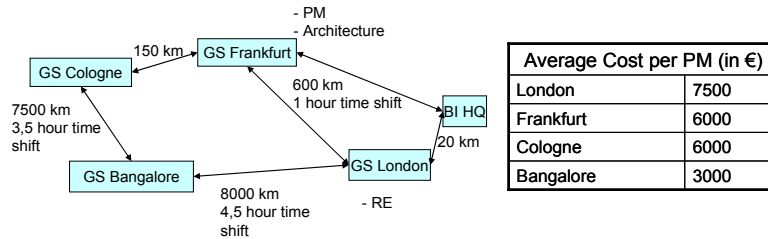

**Fig. 2.** Available Sites

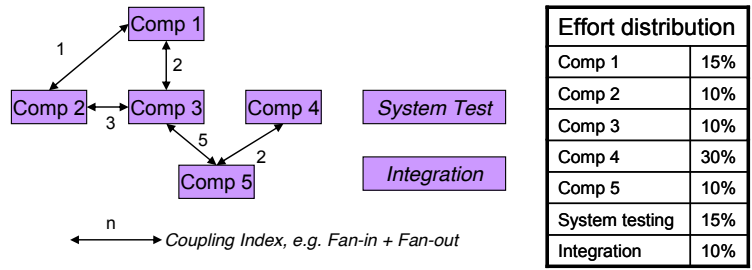

**Fig. 3.** Tasks together with the coupling between components

The problem faced by the project management is now to find an assignment of these tasks to sites that fits best with the concrete project environment and goals. This means that it must be possible to evaluate and compare the assignment of certain tasks to India versus the traditional assignment of all work to the sites in Europe. This decision has to be made with respect to a large set of influencing factors (such as the expertise available at the sites and their physical and cultural distance).

## 3. Related Work

In the following section, existing models and approaches for selecting and evaluating alternatives are briefly introduced, following by a short discussion of the applicability of each approach.

Approaches for decision support in task allocation can be classified into two groups: Optimization approaches that aim at identifying the best task assignment with respect to some goal function and predictive approaches that try to assess one specific assignment and thus can help to evaluate different assignment alternatives.

Mockus and Weiss [9] present an optimization algorithm for assigning work (chunks of modules) to sites that minimizes the communication need between the sites and thus minimizes the inherent overhead of distributed development. This model is clearly defined and easily applicable, but it focuses on only one criterion. In the given scenario, this is only of limited use, as it would neglect many important influencing factors such as the capability or cost rate at the different sites.

Another optimization approach was presented by Lamersdorf et al. [4]. In contrast to the previous approach, the focus here was placed on supporting various and conflicting criteria and on finding task assignment suggestions under conditions of uncertainty. However, this approach focuses much on the algorithmic identification of task assignments. A standard causal model of influencing factors and their impact on the decision was derived empirically but a process for customizing the model for a specific organization has not yet been defined.

Evaluation models focus on the influencing factors and their impact on one specific assignment rather than on the algorithmic identification of best assignments. Often, this is done by extending the COCOMO approach for effort estimation. Madachy [10], for instance, developed an extension of COCOMO that is able to describe site-dependent effort multipliers and thus model the impact of assigning work to specific sites. The effort multipliers, however, do not reflect the impact of distributed collaboration (e.g., physical, time-zone, or cultural distance).

Keil et al. [11] address this issue by suggesting a set of new effort multipliers that explicitly consider the impact of distributed collaboration. But this approach only names a set of multipliers without justification and also does not quantify the impact.

Another evaluation model is presented by Sooraj and Mohapatra [12], who developed an index for comparing different assignments with respect to multiple criteria. The model is based on empirical studies, but there is no explanation of how to customize the model to a specific environment and use it in this specific environment.

In summary, the existing approaches do not or only insufficiently take into account that an evaluation model has to be tailored to a specific organization and thus do not address the problem of developing a company-specific evaluation approach.

## 4. An Approach for Systematic Task Allocation Evaluation

In this section, we present an approach for systematically evaluating task allocation alternatives. It is based on the GQM approach for systematic and goal-driven measurement and evaluation and contains a set of logical steps that need to be performed for evaluating task assignments. In the following, an overview on the approach will be given first, followed by a description of each logical step.

*Approach Overview*. When deciding on the allocation of tasks to sites within a GSD project, decision makers are confronted with the problem of evaluating task allocations: Based on the evaluation and comparison of possible alternatives, their task is to identify the assignment which is most suitable for the specific project situation. The problem of task allocation can thus be reduced to finding a systematic evaluation of task alternatives with respect to the project goals and the project constraints.

The approach presented here aims at highlighting the steps that have to be performed in order to arrive at a systematic evaluation. Particularly, the factors influencing a decision and their relative weight have to be determined individually for every project. The goals of the approach can thus be described as follows:

**Goal 1**: Identify the project-specific influencing factors for a task allocation decision and their impact.

**Goal 2**: Evaluate the possible task allocation alternatives according to the project-specific influencing factors.

The approach is formulated as a set of logical steps that are based on GQM. Particularly, the GQM abstraction sheet [7] is used as a means for finding the relevant influencing factors for a specific evaluation goal. An overview of the abstraction sheet and the related process steps is given in Figure 4.

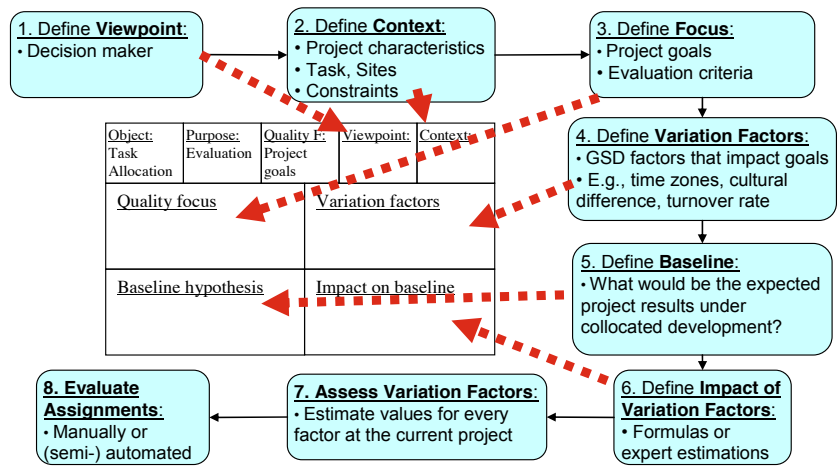

**Fig. 4.** GQM abstraction sheet together with process overview

The evaluation, specified according to the GQM goal template [7], is: "**Analyze** the task allocation **for the purpose** of evaluation **with respect to** the project goals **from the viewpoint of** […] **in the context of** […]". Based on this goal and the project-specific context and viewpoint, our approach aims at identifying the measures for project success, a baseline for the evaluation, relevant influencing factors of distributed development, and the impact of the variation factors on project success.

Depending on the maturity of the organization and the availability of data, the process steps can be supported by repositories and/or decision models. For example, a company might possess an experience base of organization-wide influence factors together with data on their impact. In this case, the project-specific selection of influencing factors would consist of selecting the relevant factors from the organizational repository.

*Process Steps.* In the following, we will list all steps of the evaluation process together with its application in the previously defined scenario.

**1. Define Viewpoint**: At first, the viewpoint of the decision (i.e., the decision maker) must be identified. This is the person responsible for the decision and the one who has the relevant information about the context of the allocation decision.

*Scenario*: At GlobalSoft, the task allocation decision is made by the responsible project manager. As he was also in charge of previous projects with BigIndustries, he knows the project and the involved participants very well.

**2. Define Context**: The context of the project comprises the characterization of the available sites (and their relations such as time-zone distances), the distribution of work to different tasks, and constraints on the task allocation decision (i.e., already assigned tasks). It thus defines the input for the task allocation decision.

*Scenario*: The context of the project, the tasks, and the sites were already described in Section 2.

**3. Define Focus**: The focus of the evaluation is on the project goals. Now, these goals have to be specified further: Which criteria define project success (e.g., cost, quality)? The different possible assignments will later be evaluated with respect to these criteria. If possible, the measures should be quantifiable. If different measures are named, a strategy for weighting them against each other should also be defined (e.g., if assignment A results in higher cost and higher quality compared to assignment B, which is rated as being suited better with respect to the goals and the given context?).

*Scenario*: In order to simplify the scenario, the total development costs are selected as the only criterion in the quality focus: The assignment with the lowest expected development costs is to be selected. However, in contrast to many approaches in practice, hourly cost rates are not the only issue that is regarded. Instead, the evaluation focus is on a realistic estimation of development costs, which also includes an estimation of the individual productivity at each site (which is determined by both site-specific factors and the overhead due to communication across sites).

**4. Define Variation Factors**: Variation factors are all those factors that have an allocation-dependent influence on the evaluation criteria. For example, if developer experience differed between sites, then assigning more work to the experienced sites would probably decrease effort. Given that effort is an evaluation criterion, developer experience would therefore be a variation factor (because its impact on effort would be dependent on the question of which tasks are assigned to the experienced or inexperienced sites).

We categorize variation factors into (a) characteristics of sites (e.g., cost rate, experience), (b) dependencies between sites (e.g., time-zone differences), (c) characteristics of tasks (e.g., size), (d) dependencies between tasks (e.g., coupling), and (e) task-site dependencies (e.g., the knowledge for performing task X existing at site Y).

*Scenario*: Based on the COCOMO II [13] effort multipliers and our own expert opinions, the following variation factors were identified:
- (a) Site characteristics: Analyst capability, programmer capability, language and tool experience, personnel continuity, customer proximity
- (b) Site dependencies: Cultural difference, time-zone difference
- (c) Task characteristics: Size
- (d) Task dependencies: Coupling
- (e) Task-site dependencies: Application experience, platform experience

**5. Define Baseline**: The goal of this process step is to derive a baseline for the success measures. Depending on the overall goal (i.e., establishing distributed development vs. modifying a distributed task assignment) and available knowledge, the baseline can reflect collocated development (all work would be assigned to one site) or an already established assignment (work would be assigned as in previous projects). The baseline may, for instance, be determined by expert estimations, historical project data, or using standard prediction models.

*Scenario*: At GlobalSoft, effort is estimated using COCOMO II. For baseline estimation, all site-dependent factors are set to the optimal case. Based on known project characteristics and the COCOMO formula, the baseline effort is estimated at 172 person-months, which are then distributed across the tasks according to the effort distribution given in Figure 3.

**6. Define Impact of Variation Factors**: In this process step, the impact of every variation factor (defined in step 4) on every criterion in the focus (defined in step 3) is evaluated. This can be done with the help of expert estimations or by analyzing past projects. For example, if effort is in the evaluation focus and time-zone difference was defined as a variation factor, the step should answer the question "How does a certain time-zone difference between two sites affect the effort overhead for tasks assigned to these sites?" If possible, this should be done quantitatively.

*Scenario*: GlobalSoft chose to use the CoBRA® approach [14] for cost estimation. This method provides a way for describing a causal model with influencing factors on development effort. Figure 5 shows the derived causal model. The quantification of the impact was done by experienced project managers at GlobalSoft. As the complete model is quantified, it is implemented in MS Excel.

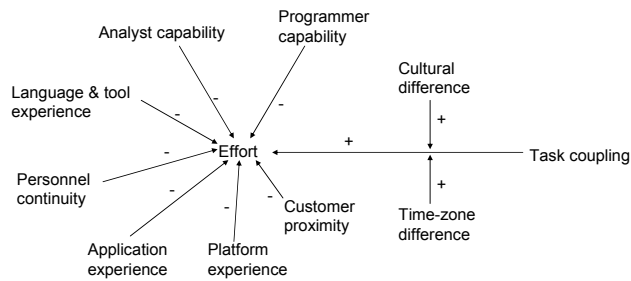

**Fig. 5.** Causal model for impact of variation factors

**7. Assess Variation Factors**: For all tasks and sites identified in step 2, the values of the variation factors are now assessed for the project at hand.

*Scenario*: The project manager assesses all values and inserts them into the Excel model.

**8. Evaluate Assignment Alternative**: Finally, every possible assignment can now be evaluated using the results of the previous steps. Depending on whether quality focus and impact of variation factors were described quantitatively or not, the evaluation can now provide predictions or guidelines and hints for every assignment that is of interest.

*Scenario*: The project manager investigates three alternatives: Assigning all work within Europe, assigning system testing and component 4 to India, and assigning everything to India. He is now able to evaluate all of them. The results show (Table 4) that assigning all work within Europe would lead to the lowest effort. Assigning parts of the work to India leads to the lowest development costs (but the difference is not very large due to the decrease in productivity). However, assigning all work to India would again increase the total costs because of the large additional effort. Based on the results, it is decided to assign component 4 and system testing to India.

Table 1. Result of the assessment: Impact on effort (person-months) and cost (in 1000 euros)

|  | Comp 1 | | Comp 2 | | Comp 3 | | Comp 4 | | Comp 5 | | System Test | | Integration | | **Total** | |
| --- | --- | --- | --- | --- | --- | --- | --- | --- | --- | --- | --- | --- | --- | --- | --- | --- |
|  | PM | Cost | PM | Cost | PM | Cost | PM | Cost | PM | Cost | PM | Cost | PM | Cost | **PM** | **Cost** |
| All in Europe | 75 | 451 | 40 | 237 | 55 | 328 | 176 | 1058 | 43 | 258 | 84 | 626 | 38 | 283 | **510** | **3241** |
| Comp 4, Testing: India | 77 | 464 | 41 | 243 | 56 | 337 | 272 | 816 | 49 | 292 | 179 | 536 | 40 | 299 | **713** | **2987** |
| All in India | 147 | 440 | 109 | 328 | 131 | 393 | 226 | 679 | 113 | 338 | 146 | 437 | 136 | 408 | **1007** | **3022** |

## 5. Conclusion and Future Work

In this article, we presented a series of logical steps that have to be performed in order to systematically evaluate possible task allocation alternatives. The steps are described on a very high level and thus have to be instantiated individually for every organization. However, as they focus on the development of an organization-specific approach for task allocation evaluation and selection, they go beyond the approaches in the literature, which typically present a rigid model for task allocation without dealing with adaptation to company-specific needs.

The instantiation of the process was done in a very optimistic scenario: All necessary information about the relevant influencing factors and their impact was available in a quantitative form, which made it possible to develop a specific quantitative model and thus exactly quantify the results of every task allocation alternative. The selection of total development costs as the only evaluation criterion also increased the ability to quantify the model (but might not be realistic in industry). In reality, however, the available knowledge is not always as detailed and quantifiable as shown here. In this case, the process steps have to be instantiated in a different way, for example by formulating qualitative rules or guidelines on the impact of certain factors.

Another limitation of the approach is that it assumes a relatively high degree of freedom in the task allocation decision regarding a specific project. In reality, however, the decision is often predefined due to specializations (especially in standard software development) and long-term strategic goals of higher management. Still, in many cases a systematic evaluation of alternatives (as presented here) promises to result in higher project success than unsubstantiated task allocation decisions focusing on cost rates only (while neglecting the impact of distribution on productivity).

In future work, we plan to apply the process steps to a real industrial global software development project in order to evaluate the approach. Based on the results, future work will also have to develop a more detailed process for evaluation and support it with tools and experience bases.

As discussed in Section 3, task allocation decision support can be seen as support for evaluating alternatives (as presented here) or as algorithmic identification of assignments. If the set of possible assignments grows over a certain size, it might not be practical to evaluate all assignments manually. We developed TAMRI, a model and tool for providing decision support in this case [4]. In future work, we plan to combine the two approaches by developing a method for project-specific task allocation evaluation and, based upon it, an algorithm for suggesting possible task assignments.